\documentstyle[aps]{revtex}

\draft

\begin{document}
\title{Neutrino oscillation mechanism for pulsar kicks reexamined }
\author{M. Barkovich$^{\dagger }$, J. C. D'Olivo$^{\dagger }$,
R. Montemayor$^{\ddag} $, J.F. Zanella$^{\ddag }\thanks{
Present Address: Dpto. de F\'{\i}sica, FCEN, Universidad de Buenos Aires,
Argentina}$}
\address{$^{\ddag }$Centro At\'{o}mico Bariloche and Instituto Balseiro,
CNEA and Universidad Nacional de Cuyo,\\
8400 S. C. de Bariloche, R\'{\i}o Negro, Argentina\\
$^{\dagger }$Departamento de F\'{\i}sica de Altas Energ\'{\i}as,
Instituto de Ciencias Nucleares,\\
Universidad Nacional Aut\'{o}noma de M\'{e}xico,\\
Apartado Postal 70-543, 04510 M\'{e}xico, Distrito Federal,
M\'exico}
\maketitle

\begin{abstract}
Anisotropic neutrino emission during the neutron star formation
can be the origin of the observed proper motions of pulsars. We
derive a general expression for the momentum asymmetry in terms
of the neutrino energy flux gradient, and show that a
nonvanishing effect is induced at the lowest order by a deformed
neutrinosphere. In particular, this result is valid for a
neutrino flux transported through a spherical atmosphere with
constant luminosity.
\end{abstract}


\section{Introduction}

Observations show that pulsars have peculiar proper motions. They have very
high translational velocities with respect to the surrounding stars, with a
mean value of $450$ $km/s$ and up to a maximum of about $1000$ $km/s$\cite
{velocidad}. This suggests that some kind of impulse (kick) happens during
the birth of the neutron star. Different mechanisms have been proposed to
explain the kick, but most of them have difficulties to produce the large
observed velocities.

Neutrinos carry away almost all the energy released in the gravitational
collapse ($\approx 3\times 10^{53}\;erg$), taking with them a momentum $\sim
100$ times the momentum associated with the spatial motion of pulsars.
Therefore, a 1\% anisotropy in the momentum distribution of the outgoing
neutrinos would suffice to account for the translational kick.

An interesting mechanism to explain the asymmetric neutrino emission from a
cooling protoneutron star has been proposed by Kusenko and Segr\`{e} \cite
{kusegre}. It is based on the matter neutrino oscillations in the presence
of an intense magnetic field. The emission surface of the electron neutrino
is located at a radius larger than the one corresponding to the muon or tau
neutrino. Under suitable conditions, a resonant transformation $\nu
_{e}\rightarrow \nu _{\tau }$ can take place in the region between the
boundaries of the electron and the tau neutrinospheres. The $\nu _{e}$ are
trapped by the medium, but the $\nu _{\tau }$ produced in this way are
outside their neutrinosphere and free to escape from the protostar.
Consequently, the surface of resonance acts as an effective tau
neutrinosphere. If there is a magnetic field, or another non-isotropic
effect, this surface of resonance becomes distorted and an anisotropy in the
energy flux is generated, causing a kick to the protostar.

Doubts about the effectiveness of the above mechanism have been
raised by Janka and Raffelt\cite{jr}. According to them, no
effect is generated at lowest order because it is not justified to
calculate the flux asymmetry from the temperature variation
around the surface of resonance. The neutrino luminosity in the
protoneutron star is controlled by the core emission and is not
affected by local processes in the atmosphere, where the flavor
transformation occurs. In support of their argument, Janka and
Raffelt use the Eddington model for a plane-parallel stellar
atmosphere \cite{ss} to estimate a residual asymmetry that is
induced by higher-order corrections.

In this work we reconsider the problem of the neutrino
oscillation mechanism for pulsar kicks. In fact, a neutron
protostar is a very complex system, mainly formed by interacting
nucleons, electrons, photons and neutrinos, rotating quickly and
with a strong magnetic field. However, for our analysis a
simplified and perturbative description is suitable. The star can
be considered as constituted by two gases, the nucleon gas and
the neutrino gas. From the hydrodynamic point of view electrons
and photons can be ignored. The dynamics of these gases are
controlled by the gravitational field, the magnetic field, the
neutrino-nucleon interactions and the hydrostatic equilibrium
equations. At the lowest order in this perturbative approach, and
neglecting the star rotation, the dynamics of the nucleon gas is
dominated by the gravitational field and the equilibrium
hydrodynamic equations, and it is thus described by a spherical
distribution. Besides this we have the neutrino gas, interacting
with the nucleon gas and the magnetic field. This last
interaction acts in particular on the neutrino mass columns,
breaking their isotropy. This breaking alters the hydrostatic
equilibrium equations, producing a non-isotropic distortion in
the nuclear matter. In the following we neglect these
contributions, but even so there is a non vanishing kick effect
induced by the geometrical deformation of the resonance surface,
provided that it acts as an effective emission surface.

We parametrize the resonance surface as $R=R_r+\delta \cos\theta$,
and develop the expressions up to first order terms in
$\delta/R_r$. From this approach we derive an expression for the
fractional momentum asymmetry in terms of the spatial derivative
of the energy flux. Using this result, we show that the
distortion of the resonance surface by the presence of a magnetic
field generates a geometrical asymmetry in the neutrino emission,
even in the case of a constant luminosity. To illustrate this
effect, we consider two simple self-consistent models for a
spherical protostar atmosphere, which satisfy the energy flux
conservation. In particular, in one of the models we use the
Eddington approximation adapted to the spherical geometry. The
required magnetic field results to be one order of magnitude
higher than the value derived in the original
articles\cite{kusegre,kks}.

To make the comparison with the existing literature easier, we
restrict the discussion to the standard mechanism of neutrino
oscillations between massive active neutrinos, and a
magnetic-field induced deformation of the resonance surface.
Nevertheless, our approach can be straightforwardly extended to
other situations \cite{sm}, and in particular to oscillations
produced by a violation of the equivalence principle (VEP)\cite {h,cdmu,bcdm}.
In the case of VEP no magnetic field is needed to deform the resonance
surface\cite{bcdm}.

In the following section we examine the effect of the deformation of the
resonance surface on the neutrino energy flux, emphasizing the relevance of
the geometrical variation of flux to produce an asymmetric momentum
emission. In Sections III and IV, we apply the results of Section II to the
Eddington and the polytrope neutrinosphere models and estimate the magnitude
of the magnetic field required to explain the observations. The last section
presents some general conclusions.

\section{Surface of resonance and neutrino energy flux}

We consider oscillations between two neutrino flavors, say $\nu
_{e}$ and $\nu _{\tau }$, in the interior of a protostar.
Neutrinos have an average energy $E\cong k=\left| {\bf k}\right|
$, which depends on the radial coordinate $r$. In the absence of
a magnetic field, or any isotropy-breaking interaction, the
resonant transformation takes place on the surface of a sphere of
radius $R_r$, given by the condition
\begin{equation}
\frac{\Delta m^{2}}{2k_{r}}\cos 2\theta =\sqrt{2}G_{F}{N_{e}}_{r}\;,
\label{r0}
\end{equation}
where $k_{r}=k(R_{r})$ and ${N_{e}}_{r}=N_{e}(R_{r})$. Here,
$\theta $ is the vacuum mixing angle, $G_{F}$ is the Fermi
constant, and $\Delta m^{2}=m_{2}^{2}-m_{1}^{2}$ is the
difference between the square mass of the neutrinos. We assume
that the number density of electrons $N_{e}(r)$ is proportional
to the baryon density $\rho (r)$, $N_{e}(r)\simeq
\frac{0.1}{m_{n}}\rho (r)$.

In the absence of a magnetic field the emission surface is
spherical, the outgoing energy flux is radial and there is no
kick, but the presence of a magnetic field, described here by a
uniform field ${\bf B}$, distorts the surface of resonance. In the
simplest case this is now defined by a function $R(\vartheta
)\simeq R_{r}+\delta \cos \vartheta $, where $\delta < R_{r}$ and
$\vartheta $ is the angle between the vector position of a point
at the surface and the direction of the magnetic field of the
protostar. The distortion of the resonance surface leads to a
modification in the outgoing energy flux. The kick is
characterized by the fractional momentum asymmetry factor
\begin{equation}
\frac{\Delta
k}{k}=\displaystyle\frac{1}{6}\displaystyle\frac{\displaystyle
\int_{0}^{\pi }{\bf F}_{s}(\vartheta )\cdot {\bf
\hat{B}}\;da}{\displaystyle \int_{0}^{\pi }{\bf F}_{s}(\vartheta
)\cdot {\bf \hat{n}}\;da}\;,  \label{kc}
\end{equation}
where ${\bf F}_{s}$ is the outgoing electron neutrino energy flux
at the element of area $da$ of the emission surface. The integrals
in the denominator and numerator give the total momentum lost by
the protostar per unit of time and its component along the
direction of the magnetic field, respectively. The factor $1/6$
comes from the fact that only the electron neutrino contributes
to the energy flux asymmetry.

To compute the asymmetry factor it is necessary to take into
account the structure of the flux at the resonance surface,
considered as an effective emission surface. The details of this
analysis are presented in the appendix. The basic information is
given by the distribution function for neutrinos in thermal
equilibrium with the medium and satisfying the diffusion
approximation
\begin{equation}
f_{\nu }\simeq f_{\nu }^{eq}-\frac{1}{\Lambda }{\bf
\hat{\Omega}}\cdot {\bf \nabla }f_{\nu }^{eq}=f_{\nu
}^{eq}-\frac{1}{\Lambda }{\bf \hat{\Omega}} \cdot {\bf
\hat{r}}\,\frac{df_{\nu }^{eq}}{dr},  \label{df}
\end{equation}
where ${\bf k}=k{\bf \hat{\Omega}}$ is the momentum of the
neutrinos, $ f_{\nu }^{eq}=\left( 1+e^{(k-\mu _{\nu })/T}\right)
^{-1}$ is the neutrino distribution function at equilibrium, $\mu
_{\nu }$ is the chemical potential, and the factor $\Lambda $
comes from the differential cross section for neutrino reactions.
The energy flux is defined by this distribution function
according to
\begin{equation}
{\bf F}_\nu({\bf r})=\int \frac{d^{3}k}{(2\pi )^{3}}\;{\bf
k}f_{\nu }({\bf r}, {\bf k}) \;. \label{jflux}
\end{equation}

At the interior of the resonance surface the energy flux is radial
(${\bf F} =F\,{\bf \hat{r}}$) and results from the diffusive part
of the neutrino distribution. As discussed in Ref. \cite{jr}, in a
neutron protostar the neutrino luminosity $L_{c}$ is governed by
the energy loss from the core. Throughout the neutrinosphere,
within the resonance surface, the luminosity can be assumed as not
being dependent on the radial coordinate, and thus as satisfying
$F(r)=\frac{L_{c}}{4\pi r^{2}}$. Thus, below the limit surface,
the flux is purely diffusive, and is given by the constant
luminosity condition. Once the energy flux above the limit
surface is computed using the expression (\ref{jflux}), we find
that it has a radial component, associated to the diffusive part
of the neutrino distribution, and a normal component, due to the
isotropic part of the neutrino distribution
\begin{equation}
{\bf F}_{s}=F_{\hat{n}}\;{\bf \hat{n}}+F_{\hat{r}}\;{\bf \hat{r}},
\end{equation}
where ${\bf \hat{n}}$ and ${\bf \hat{r}}$ are unit vectors along
the normal and radial directions respectively, at the considered
point on the resonance surface. The local flux conservation
(${\bf \nabla }\cdot {\bf F}=0 $) implies that, at the resonance
surface, the normal outgoing flux is equal to the diffusive
outgoing flux, and that both are equal to one half of the
diffusive flux calculated just below the resonance surface
\begin{equation}
F_{\hat{r}}(\vartheta )=F_{\hat{n}}(\vartheta )=\frac{1}{2}F_\nu
(R_{r}+\delta \cos \vartheta )\simeq \frac{1}{2}F_\nu
(R_{r})\left( 1+h_{F}^{-1}\delta \,\cos \vartheta \right) \,,
\end{equation}
where $h_{F}^{-1}=\displaystyle\frac{1}{F}\left.
\frac{dF}{dr}\right| _{R_{r}}$.

In the integrals of Eq. (\ref{kc}), $da$ is the element of area
on the distorted surface of resonance and does not coincide with
the element of area $da_{r}$ on the sphere of radius $R_{r}$.
These areas are related as follows:
\begin{equation}
da=\left[ 1+\left( \frac{1}{R}\frac{dR}{d\vartheta }\right)
^{2}\right] ^{1/2}2\pi R^{2}\sin \vartheta d\vartheta d\varphi
\simeq (1+2\frac{\delta }{ R_{r}}\cos \vartheta )da_{r}\;.
\label{ap}
\end{equation}
In addition, we have ${\bf \hat{r}}\cdot {\bf \hat{B}}=\cos \vartheta $, and
\begin{equation}
{\bf \hat{n}}\cdot {\bf \hat{B}}=\left[ 1+\left(
\frac{1}{R}\frac{dR}{ d\vartheta }\right) ^{2}\right]
^{-1/2}\left[ \cos \vartheta +\frac{1}{R} \frac{dR}{d\vartheta
}\sin \vartheta \right] \simeq \cos \vartheta -\frac{ \delta
}{R_{r}}\sin ^{2}\vartheta \;.  \label{cp}
\end{equation}

From Eqs. (\ref{kc})-(\ref{cp}), to first order, we observe that
only $F_{ \hat{n}}$, the component of the energy flux normal to
the resonance surface, gives a non null contribution to the
integral in the numerator of the asymmetry factor, and it is thus
responsible for the kick. It results
\begin{equation}
\frac{\Delta k}{k}=\frac{1}{36}h_{F}^{-1}\delta \;.  \label{k}
\end{equation}
Taking into account the energy emitted by the protostar in form of
neutrinos, this ratio must have a value of the order of $10^{-2}$ to produce
the observed kicks. This expression clearly shows that the existence of a
kick requires a nonvanishing gradient of the flux, i.e. $h_{F}^{-1}\neq 0$.
Using $h_F^{-1} = -2R_r^{-1}$, we obtain
\begin{equation}
\frac{\Delta k}{k} = -\frac{1}{18}\frac{\delta }{R_{r}}\;.
\label{kick}
\end{equation}
This means that $\delta $ must be of the order of $R_{r}/6$ to
produce the required kick.

To calculate $\delta$, let us remember that the index of refraction of
the neutrinos is modified by the presence of an external magnetic field
\cite{dnp}. This fact affects the flavor transformations of mixed neutrinos
producing an anisotropic contribution to the resonance condition \cite{dnec}.
For neutrinos propagating through a degenerate electron gas, the
resonance condition becomes
\begin{equation}
\frac{\Delta m^{2}}{2k}\cos 2\theta
=\sqrt{2}G_{F}N_{e}+\frac{eG_{F}}{\sqrt{2 }}\left(
\frac{3N_{e}}{\pi ^{4}}\right) ^{1/3}B\cos \vartheta \;,
\label{rc}
\end{equation}
where $e\ $is the electron charge, $B=\left| {\bf B}\right| $,
and $k$ and $ N_{e}$ are evaluated at $R(\vartheta )$. In Eq.
(\ref{rc}) it has been assumed that the weak-field limit is
satisfied, i.e. $2eB\ll(3\pi^2 N_e)^{2/3}$. More general features
of the neutrino propagation in magnetized media, incorporating the
effect of strong magnetic fields, have been considered by several
authors \cite{sem}. In our case, the  weak-field condition means
$B\ll 5\times 10^{16} G$ and,  in fact, is marginally satisfied
in the models discussed below. However, taking into account the
ambiguities in the values considered for the parameters of a
protoneuton star, a more accurate computation is not necessary.
This could be meaningful in the context of a more precise and
detailed model, which is beyond the scope of this article.

By writing $k=k_{r}+\delta k$ and $
N_{e}={N_{e}}_{r}+\delta N_{e}$, from Eqs. (\ref{r0}) and
(\ref{rc} ) we obtain the relation
\begin{equation}
\frac{\Delta m^{2}}{2k_{r}^{2}}\;\delta k\;\cos 2\theta \simeq
-\sqrt{2} G_{F}\;\delta N_{e}-\frac{eG_{F}}{\sqrt{2}}\left(
\frac{3N_{e}}{\pi ^{4}} \right) ^{1/3}B\cos \vartheta \;.
\label{rc1}
\end{equation}
There are two contributions to $\delta N_{e}$. One is due to the geometrical
distortion of the surface of resonance, assuming that the different profiles
of the neutrinosphere remain unchanged. The other comes from the distortion
of the profiles of temperature, pressure, density, etc., induced by the
geometrical distortion. The equations that define the model give the
relation between the last higher-order contribution and the deformation of
the surface of resonance. In this work we will only consider the first order
contribution. Thus, we have
\begin{equation}
\delta N_{e}=\left. \frac{dN_{e}}{dr}\right| _{R_{r}}\delta R\equiv
h_{N_{e}}^{-1}{N_{e}}_{r}\delta \cos \vartheta \;.  \label{dn}
\end{equation}
Analogously, we also have
\begin{equation}
\delta k=\left. \frac{dk}{dr}\right| _{R_{r}}\delta R\equiv
h_{k}^{-1}k_{r}\delta \cos \vartheta \;.  \label{kn}
\end{equation}
Inserting Eq. (\ref{dn}) and (\ref{kn}) into Eq. (\ref{rc1}), we get \cite
{kks}
\begin{equation}
\;\delta \simeq -\left. \frac{e}{2}\left( \frac{3}{\pi
^{4}}\right)
^{1/3}N_{e}^{-2/3}B\frac{1}{h_{k}^{-1}+h_{N_{e}}^{-1}}\right|
_{R_{r}}\;.
\end{equation}

If we assume that the electron neutrinos are in thermal equilibrium with the
stellar medium, the average energy of the emitted neutrinos is proportional
to the temperature at the emission point, $k=\frac{7\pi ^{4}}{180\zeta (3)}
T\simeq 3.15T$. In such a case, from Eq. (\ref{r0}) we get
\begin{equation}
\Delta m^{2}\cos 2\theta \simeq \frac{G_{F}\rho _{r}T_{r}}{m_{n}}\;.
\label{rs}
\end{equation}
To have the resonance within the electron neutrinosphere, $\rho _{r}T_{r}$
must be larger than the corresponding value at the surface of the
neutrinosphere, $\rho _{\nu _{e}}T_{\nu _{e}}$. For an ideal gas this simply
means that the pressure at the resonance must be larger than the pressure at
the surface of the neutrinosphere. The $\nu _{e}$ trapping density is$\;\rho
\gtrsim 10^{11}\;g\;cm^{-3}$, and hence
\begin{equation}
\Delta m^{2}\cos 2\theta >\frac{G_{F}\rho _{\nu _{e}}T_{\nu _{e}}}{m_{n}}
\simeq 4.3\times 10^{-8}\times T_{\nu _{e}}\;,  \label{rs1}
\end{equation}
where the temperature $T_{\nu _{e}}$ is given in $MeV$. For $T_{\nu
_{e}}\simeq \left( 3-5\right) \;MeV$, it requires $m_{{\nu }_{\tau }}\gtrsim
100$\ $eV$. Now, $h_{k}=h_{T}$ and, together with $h_{N_{e}}=h_{\rho }$, we
obtain
\begin{equation}
\;\delta \simeq -\left. \frac{3eB}{2}\left( \frac{10m_{n}}{3\pi ^{2}}\right)
^{2/3}\rho ^{-2/3}\frac{1}{h_{T}^{-1}+h_{\rho }^{-1}}\right| _{R_{r}}\;.
\label{delta}
\end{equation}

In general, to compute $\delta $ and $h_{F}^{-1}$ (or $R_{r}$ in the case of
Eq. (\ref{kick})) a model for the neutrino atmosphere of a neutron protostar
must be specified. This will be done in the next sections, where we examine
two analytical models that are meaningful up to the neutrinosphere, where
the neutrino transport equation holds. In both models the neutrino transport
is in the diffusion regime and the luminosity is independent of $r$, but
they differ in the assumed properties for the medium. In the first one, the
Eddington model, the medium is an ideal gas of nucleons, while in the second
one it is a polytrope gas.

\section{The spherical Eddington model}

The Eddington model gives a simple and physically reasonable description of
a neutrino atmosphere, locally homogeneous and isotropic. For a plane
geometry the model was developed by Schinder and Shapiro \cite{ss}, and here
we extend it to the spherical geometry.

For neutrinos and antineutrinos in thermal equilibrium with the
medium, and satisfying a transport regime consistent with the
diffusion approximation, the energy density, the energy flux, and
the stress tensor of neutrinos with momentum $k $ are \cite{ss}:
\begin{eqnarray}
&&U_{k} =\frac{k^{3}}{2\pi ^{2}}f_{\nu }^{eq}\;,  \label{ed1} \\
&&{\bf F}_{k} =-\frac{1}{3\Lambda }\frac{k^{3}}{2\pi ^{2}}{\bf
\nabla }
f_{\nu }^{eq}\;,  \label{fe1} \\
&&\left( T_{k}\right) _{ij} =\frac{1}{3}\delta _{ij}U_{k}\;.
\label{st1}
\end{eqnarray}
We assume that $\Lambda =\kappa \rho k^{2}$, with $\kappa
=5.6\times 10^{-9}\;erg^{-3}cm^{4}s^{-2}$. In the interior of the
neutrinosphere we can consider that there are two perfect fluids.
One of them is constituted by nonrelativistic nucleons of mass
$m_{n}$, with a density $\rho $, and the other by
ultrarelativistic neutrinos and antineutrinos, with a vanishing
chemical potential, $\mu _{\nu }=\mu _{\bar{\nu}}=0$. Photons and
electrons are of course present and, in fact, electrons make the
relevant contribution to the effective potential in the case of
matter oscillations between active neutrinos. However, we can
ignore both of them for the hydrodynamic description of the
system. Therefore, the relationships between pressure $P$ ,
density of energy $\varepsilon $, and temperature $T$ are
\begin{eqnarray}
&&P =\frac{\rho }{m_{n}}T+\frac{7\sigma }{24}T^{4}\;, \\
&&\varepsilon =\rho +\frac{7\sigma }{8}T^{4}\;,
\end{eqnarray}
where $\sigma \simeq 2.09\times 10^{49}\;erg^{-3}cm^{-3}$ is the
Stefan-Boltzmann constant. Taking into account the gravitational field $\phi
$ of the star, we have for $P$
\begin{equation}
{\bf \nabla }P=-(P+\rho ){\bf \nabla }\phi \;.
\end{equation}

Putting all these relations together, we obtain the set of equations which
describes an isotropic neutrinosphere

\begin{eqnarray}
&&U(r) =\frac{7}{8}\sigma T^{4}(r)\;,  \label{m1} \\
&&F(r) =-\frac{1}{36}\frac{1}{\kappa \rho (r)}\frac{d}{dr}T^{2}(r)\;,
\label{m2} \\
&&P(r) =\frac{\rho }{m_{n}}T(r)+\frac{7\sigma}{24}T^{4}(r)\;,  \label{m3} \\
&&\frac{d}{dr}P(r) =-\left[ P(r)+\rho (r)+\frac{7\sigma }{8}T^{4}(r)\right]
\frac{GM(r)}{r^{2}}\;,  \label{m4}
\end{eqnarray}
with $M(r)=4\pi \int_{0}^{r}d\tilde{r}\;\tilde{r}^{2}\rho _{_{T}}(\tilde{r})$
, where $\rho _{_{T}}$ is the total density of mass.

In the region where a resonant transformation could happen the baryon
density is $\rho \simeq (10^{11}-10^{12})\;g\;cm^{-3}$ and $T\simeq
(3-10)\;MeV$. Thus, we have $\rho _{\nu _{e}}=\frac{7\sigma }{8}T^{4}\simeq
\left( 10^{-5}-10^{-2}\right) \rho $ and $P\simeq \left(
10^{-3}-10^{-2}\right) \rho $. Therefore, Eqs. (\ref{m3}) and (\ref{m4})
reduce to
\begin{eqnarray}
&&P=\frac{\rho T}{m_{n}}\;,  \label{mm1} \\
&&\frac{dP}{dr}=-\rho \frac{GM(r)}{r^{2}}\;.  \label{mm2}
\end{eqnarray}
These last equations together with Eq. (\ref{m2}) describe an ideal gas of
nucleons at hydrostatic equilibrium, with an energy flux given by the
transport of neutrinos.

From Eqs. (\ref{mm1}) and (\ref{mm2}) we have
\begin{equation}
\left. h_{\rho }^{-1}+h_{T}^{-1}\right| _{r=R_{r}}=-\frac{Gm_{n}M_{r}}{
R_{r}^{2}T_{r}}\;,  \label{ht}
\end{equation}
where $M_{r}$ is the protostar mass enclosed by the resonance sphere.
Therefore, from Eqs. (\ref{ht}), (\ref{delta}), and (\ref{kick}), the
necessary magnetic field results
\begin{equation}
B\simeq \left( \frac{10^{4}m_{n}}{3\pi ^{2}\rho _{r}}\right)
^{-2/3}\frac{ 12Gm_{n}M_{r}}{eR_{r}T_{r}}\;.  \label{BE}
\end{equation}
At this point, assuming that the resonance occurs near the
surface of the neutrinosphere, i.e. $R_{r}\simeq R_{\nu _{e}}$,
we can make an estimation of the order of magnitude of $B$ for a
typical protoneutron star. Then, taking $R_{r}\simeq 30\;km$,
$M_{r}\simeq M_{\odot }$, $T_{r}\simeq 5\;MeV$, and $\rho
_{r}\simeq 10^{11}\;g\;cm^{-3}$, to have $\Delta k/k\simeq 0.01$
the magnetic field must be $B\simeq 10^{16}\;G$ . Of course,
these parameters are not independent and a more careful
discussion of the model is necessary.

The neutrinosphere is defined by four functions: pressure $P(r)$,
temperature $T(r)$, baryonic density $\rho (r)$, and energy flux $F(r)$. Up
to this point we have only three independent equations relating these
functions. A complete specification of the system requires a fourth
relation. A simple analytical model that satisfies the requirement of a
constant luminosity is the Eddington atmosphere \cite{ss}. This model is
defined by Eqs. (\ref{m2}), (\ref{mm1}), and (\ref{mm2}), plus the
hypothesis of the energy flux conservation $\bigtriangledown .{\bf F}=0$.
For an isotropic flux, the additional assumption leads to:
\begin{equation}
\frac{\partial (r^{2}F)}{\partial r}=0\;,  \label{h44}
\end{equation}
which simply means
\begin{equation}
F=\frac{L_{c}}{4\pi r^{2}}\;,  \label{ff}
\end{equation}
where $L_{c}$ is the luminosity of the protostar.

From Eqs. (\ref{mm2}) and (\ref{ff}) we have for the mass density
\begin{equation}
\rho =-\frac{2\pi }{9}\frac{r^{2}}{\kappa L_{c}}T\frac{dT}{dr}\;,  \label{r}
\end{equation}
and replacing this expression in Eq. (\ref{mm2}) we arrive at
\begin{equation}
\frac{dP}{dr}=\frac{2\pi }{9}\frac{G}{\kappa L_{c}}M(r)T\frac{dT}{dr}\;.
\label{dp}
\end{equation}

To find the solution of the structure equations for this model we use the
following procedure. First we define a reduced effective mass $m(r)$, given
by
\begin{equation}
\int_{R_{c}}^{r}dr\;M(r)T\frac{dT}{dr}\equiv M_{c}m(r)\int_{R_{c}}^{r}dr\;T
\frac{dT}{dr}\;,
\end{equation}
where $M_{c}$ is the mass of the core. In terms of $m(r)$ the solution of
Eq. (\ref{dp}) is immediate. Defining $\alpha _{c}=\frac{\pi }{9}\frac{
GM_{c}m_{n}T_{c}}{\kappa L_{c}\rho _{c}}$, we have
\begin{equation}
P(r)=\frac{P_{c}}{1-a}\left( \frac{T^{2}}{T_{c}^{2}}-a\right) \;,  \label{p}
\end{equation}
where
\begin{equation}
a(r)=1-\frac{1}{\alpha _{c}m(r)}\;.  \label{am}
\end{equation}
From Eqs. (\ref{mm1}) and (\ref{p}) we can express the density in terms of
the temperature as follows
\begin{equation}
\rho (r)=\frac{\rho _{c}}{1-a}\frac{T_{c}}{T}\left( \frac{T^{2}}{T_{c}^{2}}
-a\right) \;,  \label{rr1}
\end{equation}
and using this result in Eq. (\ref{r}) we obtain a first order differential
equation for $T$
\begin{equation}
\frac{dT}{dr}+\frac{\lambda _{c}T_{c}}{1-a}\frac{R_{c}}{r^{2}}\left( \frac{
T^{2}-T_{c}^{2}a}{T^{2}}\right) =0\;,  \label{dif}
\end{equation}
where $\lambda _{c}=\frac{9}{2\pi }\frac{\kappa L_{c}\rho _{c}}{
T_{c}^{2}R_{c}}$. According to this equation the slope of the temperature at
$R_{c}$ is independent of the function $a(r)$, $T_{c}^{\prime }=-\frac{
\lambda _{c}T_{c}}{R_{c}}$. If we refer to an idealized neutrinosphere,
where this model would apply in the whole space, the interesting solutions
correspond to infinite protostars where the temperature has an asymptotic
behavior for $r\gg R_{c}$, such that the temperature tends to $T_{s}\simeq
\sqrt{a}T_{c}$. Thus, the function $a(r)$ varies in the range $1-\alpha
_{c}^{-1}<a<\left( T_{s}/T_{c}\right) ^{2}$ for $R_{c}<r<\infty $.

This system of equations has no analytical solution when $a$ is a function
of $r$, and in general there is no perturbative expansion that gives a good
approximate solution at every point within the neutrinosphere. To find an
approximate solution let us consider the differential equation (\ref{dif})
with $a$ constant. In this case, for $T_{c}>T>\sqrt{a}T_{c}$ an analytical
(implicit) solution is given by
\begin{equation}
T-T_{c}+\frac{T_{c}\sqrt{a}}{2}\left[ \ln \left(
\frac{T-T_{c}\sqrt{a}}{ T+T_{c}\sqrt{a}}\right) -\ln \left(
\frac{1-\sqrt{a}}{1+\sqrt{a}}\right) \right] =\frac{T_{c}\lambda
_{c}}{1-a}\left( \frac{R_{c}}{r}-1\right) \;. \label{tem}
\end{equation}

If we replace the constant $a$ for a (well behaved) function of $r$, then
the above expression still satisfies Eq. (\ref{dif}) at $r=R_{c}$. For an
infinite atmosphere, a good approximation to the exact solution is given by
Eq. (\ref{tem}), with $a$ now a function of $T(r)$:
\begin{equation}
a(T)=1-\frac{1}{\alpha _{c}}-A\left( 1-\frac{T}{T_{c}}\right) \,\,,
\label{at}
\end{equation}
where $A=\frac{T_{c}}{T_{c}-T_{s}}\left( 1-\frac{1}{\alpha _{c}}-\frac{
T_{s}^{2}}{T_{c}^{2}}\right) $, and $T_{s}$ denotes the temperature at the
surface of the electron neutrinosphere, which is assumed to lie in the
asymptotic region. From Eq. (\ref{at}), we see that $a(T_{s})=
\left(T_{s}/T_{c}\right)^{2}$ and $a(T_{c})=1-\alpha _{c}^{-1}$, and thus it
fits the extreme values of $a(r)$.

The surface of the electron neutrinosphere corresponds to a density $\rho
_{s}\simeq \rho _{\nu _{e}}$. Assuming that the temperature at this surface
is close to $T_{s}$, from Eq. (\ref{rr1}) we obtain

\begin{equation}
T_{\nu _{e}}\simeq T_{s}\left( 1+\frac{T_{c}}{2T_{s}-AT_{c}}\frac{\rho _{\nu
_{e}}}{\rho _{c}}\right) \;,  \label{tn}
\end{equation}
with $\rho _{\nu _{e}}\ll \rho _{c}$. The radius of the neutrinosphere is
obtained by evaluating Eq. (\ref{tem}) in $T_{\nu _{e}}$.

To illustrate the predictions of the model, we adopt a neutron
protostar with reasonable values for its parameters. The core is
assumed to have $ M_{c}=M_{\odot }=1.13\allowbreak \times
10^{60}\;MeV$, $R_{c}=10\ km$, $ L_{c}=9.5\times
10^{51}\;erg\;s^{-1}$, $\rho _{c}=10^{14}\;g\ cm^{-3}$, and $
T_{c}=40\ MeV$. We take the surface of resonance as defined by a
density $ \rho _{r}=10^{11}g\;cm^{-3}$, while a numerical
estimation for the asymptotic value of the solution of Eq.
(\ref{dif}) gives $T_{s}=4.8\ MeV$. Inserting these values into
Eqs. (\ref{tem}) and (\ref{tn}) we get $R_{\nu _{e}}\simeq
2.7R_{c}$, and a numerical computation of the total mass of the
star from Eq. (\ref{rr1}) gives $M_{s}=1.4M_{\odot }$. Then, for
a resonance region lying near the surface of the electron
neutrinosphere, from Eqs. (\ref {dif}), (\ref{rr1}), and
(\ref{at}) we have
\begin{eqnarray}
\left. h_{T}^{-1}\right| _{r=R_{r}} &\simeq &-\lambda _{c}\frac{
T_{c}^{2}R_{c}\rho _{r}}{T_{r}^{2}R_{r}^{2}\rho _{c}}\sim
-.04\frac{1}{R_{r}}
\;, \\
\left. h_{\rho }^{-1}\right| _{r=R_{r}} &\simeq &\frac{T_{r}\rho _{c}}{
T_{c}\rho _{r}}\left( 2-A\frac{T_{c}}{T_{r}}\right) \left. h_{T}^{-1}\right|
_{R_{\nu _{e}}}\sim -19\frac{1}{R_{r}}\;.
\end{eqnarray}
According to Eqs. (\ref{delta}) and (\ref{kick}), we see that a
$\Delta k/k$ of order $0.01$ can be obtained with $B\simeq
3\times 10^{16}\;G$. This value is in agreement with the
estimation done by means of Eq. (\ref{BE}). In general, for more
extended and hotter Eddington protostars smaller magnetic fields
are needed, as can also be seen from Eq. (\ref{BE}).

The strength of the magnetic field we have obtained is somewhat higher than
those estimated in previous works on the subject \cite{kusegre,kks}.
However, it is important to note that for us $B$ is at least an order of
magnitude lower than the one given in Ref.\cite{jr}. The discrepancy can be
easily understood. We have used a more realistic spherical model, where the
flux varies as $r^{-2}$ (Eq. (\ref{ff})), while in Ref. \cite{jr} the
resonance region was described in terms of a plane Eddington atmosphere,
where $F$ is constant. In the last case $h_{F}^{-1}$ vanishes and no kick is
generated at the lowest order.

\section{The polytrope model}

The conditions at the inner core of the protostar are consistent with a
polytrope gas of relativistic nucleons with an adiabatic index $\Gamma =4/3$
\cite{gw,r,st}. In this section we assume that this value of $\Gamma$ also
holds in the rest of the star. This model satisfies the same equations of
hydrodynamic equilibrium (\ref{mm2}), energy transport (\ref{m2}) and flux
conservation (\ref{ff}) as the Eddington model, but differs in the equation
of state. It leads to a confined atmosphere, where the density becomes zero
at a radius of the order of a few core radii.

The equation of state for a polytrope gas of adiabatic index $\Gamma $ is
given by \cite{weinberg,st,mh}
\begin{equation}
P=K\rho ^{\Gamma }\;.  \label{p1}
\end{equation}

From Eqs. (\ref{p1}) and (\ref{mm2}), the density profile is determined by
the integro-differential equation
\begin{equation}
\frac{d\rho ^{\Gamma -1}}{dr}=-\frac{\lambda _{\Gamma }R_{c}\rho
_{c}^{\Gamma -1}}{M_{c}}\frac{M(r)}{r^{2}}\;,  \label{PE}
\end{equation}
where $\lambda _{\Gamma }=\frac{GM_{c}}{R_{c}\rho _{c}^{\Gamma
-1}}\frac{ \left( \Gamma -1\right) }{K\Gamma }$. An extremely
good approximation for $ \rho $ is given by the function
\begin{equation}
\rho ^{\Gamma -1}(r)=\rho _{c}^{\Gamma -1}\left[ \lambda _{\Gamma }\left(
\frac{R_{c}}{r}-1\right) m(r)+1\right] \;,  \label{rg}
\end{equation}
with $m(r)=\mu +\left( 1-\mu \right) \frac{R_{c}}{r}$, such that $m(R_{c})=1$
. The radius of the star, defined by the surface where the density becomes
zero, $\rho (R_{s})=0$, determines the value of the $\mu $ parameter. The
relationship between $\mu $ and $R_{s}$ is given by
\begin{equation}
\mu =\left[ \frac{R_{s}}{\lambda _{\Gamma }\left( R_{s}-R_{c}\right) }-\frac{
R_{c}}{R_{s}}\right] \frac{R_{s}}{\left( R_{s}-R_{c}\right) }.
\end{equation}
It is convenient to rewrite Eq. ({\ref{rg}) as follows
\begin{equation}
\rho ^{\Gamma -1}(r)=\rho _{c}^{\Gamma -1}\left[ a\left( \frac{R_{c}}{r}
\right) ^{2}+b\frac{R_{c}}{r}+c\right] \;,  \label{rg2}
\end{equation}
with $a=(1-\mu )\lambda _{\Gamma}$, $b=(2\mu
-1)\lambda_{\Gamma}$, and $ c=1-\mu \lambda _{\Gamma}$}.

Once the mass density is known, the temperature profile is determined by the
condition of constant luminosity (\ref{ff}), together with (\ref{m2}):
\begin{equation}
\frac{dT^{2}}{dr}=-\frac{9\kappa L_{c}}{\pi r^{2}}\rho \;,  \label{tg}
\end{equation}
with $\rho $ given by Eq. (\ref{rg}). The solution to Eq.
(\ref{tg}) for $ \Gamma =4/3$ can be found exactly. We write it as
\begin{equation}
T(r)=T_{c}\sqrt{\lambda _{c}\left[ \chi \left( \frac{R_{c}}{r}\right) -\chi
(1)+1\right] }\;,  \label{temp}
\end{equation}
where $\chi (x)$ is a polynomial of degree seven in the variable $x$
\begin{eqnarray}
\chi (x) &=&c^{3}x+\frac{3}{2}bc^{2}\; x^{2}+c(ac+b^{2})x^{3}+\frac{b}{4}
(6ac+b^{2})x^{4}  \nonumber \\
&&+\frac{3a}{5}(ac+b^{2})x^{5}+ \frac{ba^{2}}{2}x^{6}+\frac{a^{3}}{7}x^{7},
\end{eqnarray}
where $a$, $b$, and $c$ are the parameters introduced in Eq. (\ref{rg2}).

The inverse characteristic lengths of the temperature and density at the
resonance can now be calculated from Eqs. (\ref{temp}) and (\ref{rg}),
respectively
\begin{eqnarray}
h_{T}^{-1} &=&-\lambda _{c}\frac{\rho _{r}}{\rho _{c}}\left(
\frac{T_{c}}{
T_{r}}\right) ^{2}\frac{R_{c}}{R_{r}^{2}}\;, \\
h_{\rho }^{-1} &=&-3\left( \frac{\rho _{c} }{\rho _{r}}\right) ^{1/3}\left(
2a+b \frac{R_c}{R_r}\right) \frac{R_{c}^{2}}{R_{r}^{3}}\;.  \label{hpoli}
\end{eqnarray}

To estimate the magnitude of $B$ we use the same values for the
core parameters as in the Eddington model. The constant $K$ is
fixed by the condition $K=P/{\rho }^{\Gamma }$, which gives
$K=T_{c}/m_{n}{\rho _{c}} ^{1/3}=5.6\times 10^{-5}\;MeV^{-4/3}$.
Similarly to the previous section, we adopt here $R_{s}=5.8R_{c}$
from a numerical estimation. The radius of the surface of
resonance can be calculated from Eq. (\ref{rg2}) and the result
is $R_{r}=4.2R_{c}$. Using these values in Eq. (\ref{hpoli}) we
obtain $ h_{T}^{-1}=-4\times 10^{-3}R_{r}^{-1}$ and $h_{\rho
}^{-1}=-11R_{r}^{-1}$, which substituted in Eqs. (\ref{delta})
and (\ref{kick}) yield $B\simeq 5\times 10^{16}\;G$ in order to
have $\Delta k/k\simeq 0.01$. This result is in agreement with the
values for the magnitude of the magnetic field calculated in the
previous section.

\section{Conclusions}

In this paper we revisit the resonant neutrino conversion for
pulsar kicks. By expressing the momentum asymmetry in terms of
the logarithmic derivative of the energy flux, we make clear that
a kick is produced at order $\delta/R_r$ by the radial dependence
of this flux combined with a deformation of the resonance
surface. An important ingredient for this result is the existence
of a component of the neutrino flux normal to the resonance
surface, which acts an effective tau neutrinosphere. This is valid
even though the neutrino luminosities are controlled by the core
emission. In the particular case of a plane atmosphere with
constant flux, $h_{F}^{-1}$ vanishes and there is no kick effect
to lowest order. However, with a more realistic spherical
geometry for the atmosphere this is not true anymore.

To estimate the neutrino flux anisotropy, we consider two simple
self-consistent models for the stellar atmosphere: the spherical
Eddington model and a model for a star composed of a polytropic
neutron gas. Both models take into account the energy flux
conservation and give reasonable profiles for temperature,
density, and pressure. The results clearly show that the main
effect is due to the geometrical deformation of the neutrinosphere
surface, and not to the alteration of a conserved plane-parallel
flux by a response of the global stellar structure \cite{jr}.
This last contribution is much weaker than the geometrical one.
It should be noticed that since sterile neutrinos do not interact
with the medium, the in-going $\nu_s$ produced at the resonance
surface would not be absorved. Thus, in this case, in the
approximation we use, the resonant conversion mechanism is not
able to generate a kick.

For typical values of the protoneutron star parameters, in both
models the magnetic field required to generate an appropriate
kick is of order of $B\sim 10^{16}\;G$. A more reliable
quantitative evaluation of the effect would require a detailed
model and much more involved numerical calculation. Nevertheless,
our simplified discussion indicates that the neutrino oscillation
mechanisms can not be discarded as a possible explanation for the
pulsar velocities. Here, we adopt the usual mechanism of
oscillations between massive active neutrinos, which requires a
large tau neutrino mass of the order of $100\;eV$. However, the
same analysis can be performed in other scenarios in agreement
with the present boundaries on the neutrino properties, such as
oscillations induced by a violation of the equivalence principle.

\section{Acknowledgments}

This work has been partially supported by CONICET (Argentina) and
by CONACYT (M\'{e}xico) and PAPIIT-UNAM GRANT . M. B. also
acknowledges support from SRE (M\'{e}xico).

\section*{Appendix: Energy flux on a distorted resonance surface}

Here we analyze the energy flux through the distorted surface of
resonance, a sphere with its center shifted in $\delta $ with
respect to the center of the star. At a given point on this
surface characterized by a polar angle $ \vartheta $, referred to
the direction of the magnetic field, the normal forms an angle
$\alpha $ with respect to the radial direction, given by
\begin{equation}
\sin \alpha =\frac{\delta }{R}\sin \vartheta \;.
\end{equation}
The expression for the neutrino distribution in the diffusive
approximation takes the form
\begin{equation}
f_{\nu }({\bf r},k,{\bf \hat{\Omega})}=f_{\nu }^{eq}({\bf
r},k{\bf )}+\frac{1 }{\kappa \rho k^{2}}\left| \frac{df_{\nu
}^{eq}}{dr}\right| \cos \beta \;,
\end{equation}
where $\beta $ is the angle between ${\bf k}=k{\bf \hat{\Omega}}$
and the radial direction ${\bf \hat{r}}$. We assume null chemical
potential.  Introducing local coordinates given by an azimuthal
angle $\chi $ on the tangent plane (with ${\bf \hat{\chi}}\cdot
{\bf \hat{n}}=0$), and a polar angle $\phi $ respect to the
interior normal ($-{\bf \hat{n}}$), $\beta $ can be expressed as
\begin{equation}
\cos \beta =\sin \chi \sin \phi \sin \alpha +\cos \phi \cos \alpha \;.
\end{equation}

At a given point of the resonance surface the produced tau
neutrinos have the same isotropic distribution as the parent
electron neutrinos. The in-going tau neutrinos that fall into
their own neutrinosphere are absorbed and thermalized. The
remaining tau neutrinos escape from the star and contribute to the
kick. In the following we will assume for simplicity that all the
in-going tau neutrinos are absorbed. Thus, from Eq. (\ref{jflux}),
restricting the integration to ${\bf k\cdot \hat{n}} \geq 0$, we
obtain that the flux has the normal component ($F_{\hat{n}}\,{\bf
\hat{n} }$) and the tangential one ($F_{\hat{\vartheta}}\,{\bf
\hat{\vartheta}}$) given by
\begin{eqnarray}
F_{\hat{n}} &=&\frac{7}{64}\sigma
T^{4}+\frac{1}{144}\frac{1}{\kappa \rho }
\left| \frac{dT^{2}}{dr}\right| \cos \alpha \;, \\
F_{\hat{\vartheta}} &=&\frac{1}{72}\frac{1}{\kappa \rho }\left|
\frac{dT^{2} }{dr}\right| \sin \alpha \;,
\end{eqnarray}
where all quantities are evaluated at the point of the resonance surface
defined by the angle $\vartheta $. Alternatively, we can describe this flux
in terms of a normal component and a radial one, which are
\begin{eqnarray}
F_{\hat{n}} &=&\frac{7}{64}\sigma T^{4}\;, \\
F_{\hat{r}} &=&\frac{1}{144}\frac{1}{\kappa \rho }\left|
\frac{dT^{2}}{dr} \right| \;,
\end{eqnarray}
respectively. The normal flux comes from the contribution of
$f_{\nu }^{eq}$ to the neutrino distribution function, while the
radial flux comes from the term which depends on its gradient. We
observe that the radial flux above the limit surface is one half
of the flux immediately below that surface,
\begin{equation}
F\left( R(\vartheta )-0^{+}\right) =\frac{1}{72}\frac{1}{\kappa
\rho }\left| \frac{dT^{2}}{dr}\right| \;,
\end{equation}
which is purely diffusive. Because $\cos \alpha =1+{\cal O}(\delta
^{2}/R^{2})$, the equation ${\bf \nabla }\cdot {\bf F}=0$,
integrated in an infinitesimally thin volume just above the limit
surface, implies that $F_{\hat{n}}$ is equal to $F_{\hat{r}}$.
Hence, for a given $\vartheta $ we have
\begin{equation}
F_{\hat{r}}(\vartheta )=F_{\hat{n}}(\vartheta
)=\frac{1}{2}F_\nu(R_{r}+\delta \cos \vartheta )\simeq
\frac{1}{2}F_\nu(R_{r})\left( 1+h_{F}^{-1}\delta \,\cos \vartheta
\right) \,,
\end{equation}
where $h_{F}^{-1}=-2R^{-1}$.

\end{document}